\documentstyle[aps,prl,epsf,floats]{revtex}
\begin{document}
\twocolumn[\hsize\textwidth\columnwidth\hsize\csname@twocolumnfalse%
\endcsname


\title{ Photoemission spectra of LaMnO$_3$ controlled by orbital excitations}
\author { Jeroen van den Brink,$^1$ Peter Horsch,$^1$ 
          and Andrzej M. Ole\'{s}$^{2,1}$  }
\address{ $^1$Max-Planck-Institut f\"ur Festk\"orperforschung,
          Heisenbergstrasse 1, D-70569 Stuttgart, Germany }
\address{ $^2$Institute of Physics, Jagellonian University, Reymonta 4,
          PL-30059 Krak\'ow, Poland }

\date{\today}
\maketitle

\begin{abstract}
We investigate the spectral function of a hole moving in the orbital-ordered 
ferromagnetic planes of LaMnO$_3$, and show that it depends critically on 
the type of orbital ordering. While the hole does not couple to the spin 
excitations, it interacts strongly with the excitations of $e_g$ orbitals 
(orbitons), leading to new type of quasiparticles with a dispersion on the 
orbiton energy scale and with strongly enhanced mass and reduced weight. 
Therefore we predict a large redistribution of spectral weight with respect 
to the bands found in local density approximation (LDA) or in LDA+U. 
\end{abstract}

]


Transition metal (TM) compounds are well known for their large diversity and 
richness in phenomena \cite{Ima98}. This variety of properties is not only
due to the strongly correlated nature of the electronic $3d$ states in these 
systems, often rendering them magnetic, and to the strong hybridization 
with the extended ligand valence states, but also due to the orbital 
degeneracy of the open $3d$ shells. In a localized system such orbital 
degeneracy will be lifted in one way or another -- this is the well known 
Jahn-Teller (JT) effect. In concentrated systems this often leads to 
structural phase transitions accompanied by a certain ordering of occupied 
orbitals. Alternatively, strong correlations may lead to orbital order via 
superexchange interactions \cite{Kug82}, which may dominate the JT related 
contribution \cite{Pata}. Effects of this kind are observed in TM compounds 
with three-fold $t_g$ orbital degeneracy, for example in LiVO$_2$ 
\cite{Pen97}, but are strongest in compounds with a two-fold $e_g$ orbital 
degeneracy, containing Mn$^{3+}$ or Cr$^{2+}$ ($d^4$), Co$^{2+}$ or 
Ni$^{3+}$ ($d^7$), or Cu$^{2+}$ ($d^9$) ions in octahedral coordination. 
Some well-known examples are cubic (LaMnO$_3$, KCrF$_3$ and KCuF$_3$) and 
layered (LaSrMn$_2$O$_7$ and K$_2$CuF$_4$) perovskites 
\cite{Kug82,Gehring75}.

Superexchange interactions stabilize the magnetic and orbital order 
\cite{Gooden63}, and the low-energy excitations of the system are magnons 
and orbital waves (orbitons). For simplicity we shall neglect the coupling to 
JT phonons. In fact, it has been argued by Dagotto {\it et al.} \cite{Hot99} 
that if the Hund's rule interaction $J_H$ is large, {\it the JT coupling 
leads to the same orbital pattern as superexchange\/}, and we assume that its 
effect can be absorbed in the renormalized parameters of the electronic model. 

In this Letter we present a study of single hole motion in a ferromagnetic 
(FM) orbital-ordered plane. Due to strong correlations the hole motion is
restricted by the magnetic order \cite{Brink99_1}.
In the FM and orbital ordered state \cite{Mur98} the hole can move 
without disturbing the background due to the off-diagonal hopping between 
$e_g$ orbitals, and is essentially predicted by the band structure theory 
using either the local density approximation (LDA), or its modification for 
the correlated oxides, the LDA+U method \cite{Ani91}.
The picture that arises from the present work is, however, quite different.
In addition to free propagation, the hole motion may frustrate the orbital 
order, just as it does for the antiferromagnetic order in the $t$-$J$  
model. As we show below, the interaction between the hole and the orbitals 
is so strong in the manganites that propagating holes are dressed with many 
orbital excitations and form {\it orbital polarons\/} that have large mass, 
small bandwidth and low quasiparticle (QP) weight \cite{noteeph}.

{\it Ground state\/} $-$
The ground state of LaMnO$_3$ is formed by FM layers which stagger along 
the $c$-direction, also known as A-type antiferromagnetic phase. 
As a consequence of the double-exchange \cite{Zen51} the motion along the 
$c$-axis is hindered and propagation of holes occurs only within the FM 
$(a,b)$-planes. Consider first the order of degenerate $e_g$ orbitals induced 
by superexchange in a two-dimensional (2D) FM plane with one electron per 
site ($n=1$). The orbital degrees of freedom are most conveniently described 
by pseudospins: $T_i^z=\case{1}{2}\sigma_i^z$ and
$T_i^x=\case{1}{2}\sigma_i^x$, where $\sigma_i^{\alpha}$ are Pauli 
matrices, and we label the orbital $|x\rangle\equiv |x^2-y^2\rangle$ and 
$|z\rangle\equiv |3z^2-r^2\rangle$ states by the pseudospin components: 
$\mid\uparrow\rangle\equiv |x\rangle$ and 
$\mid\downarrow\rangle\equiv |z\rangle$. In the limit of large on-site 
Coulomb interaction $U$, the effective interaction between the orbital 
pseudospins is given by the superexchange $J>0$, which involves the high-spin 
$|^6A_1\rangle$ excited states \cite{Fei99}. One finds 
\begin{equation}
H_J \!=\! \case{1}{2}J\sum_{\langle ij\rangle}
         \left[T^z_i T^z_j + 3T^x_i T^x_j
   \mp\! \sqrt{3} (T^x_i T^z_j+T^z_i T^x_j)\right],
\label{H_J}
\end{equation}
where $\langle ij\rangle$ denotes a nearest neighbor pair, and the negative 
(positive) sign applies to a bond along the $a(b)$-axis. In the ground state 
the occupied $e_g$ orbitals are staggered:
$|\mu\rangle=(|x\rangle+|z\rangle)/\sqrt{2}$ and 
$|\nu\rangle=(|x\rangle-|z\rangle)/\sqrt{2}$ on $A$ and $B$ sublatice, 
respectively \cite{notejt}. These orbitals define locally a new basis 
obtained from $\{|x\rangle,|z\rangle\}$ orbitals by a rotation by the same 
angle on the two sublattices, $\psi_A=\psi_B=\frac{\pi}{4}$ \cite{Bri99}.

We write the kinetic energy for holes introduced in the orbital-ordered
state in terms of this new orbital basis,
\begin{eqnarray}
H_t&=&\case{1}{4}t\!\sum_{\langle ij\rangle}\left[
  f_{i0}^{\dagger}f_{j0}^{}+f_{i1}^{\dagger}f_{j1}^{}
 +2(f_{i1}^{\dagger}f_{j0}^{}+f_{i0}^{\dagger}f_{j1}^{}) \right. \nonumber \\
& & \hskip .7cm \left. \pm\sqrt{3}\left(
   f_{i1}^{\dagger}f_{j0}^{}-f_{i0}^{\dagger}f_{j1}^{}\right)+H.c.\right],
\label{eq:ht01}
\end{eqnarray}
where $f_{i\alpha}^{\dagger}$ creates a hole on site $i$ in orbital $\alpha$, 
and upper (lower) sign applies to a bond $\parallel a(b)$-axis. Here the 
indices $0$ and $1$ refer to the occupied ($|0\rangle$) and unoccupied 
($|1\rangle$) orbital states, as determined by $H_J$. The hopping integrals 
in Eq. (\ref{eq:ht01}) follow from the $e_g$ symmetry and therefore depend 
strongly on the pair of orbitals at nearest-neighbor sites involved in a 
hopping process. 

\begin{figure}
      \epsfxsize=90mm
      \centerline{\epsffile{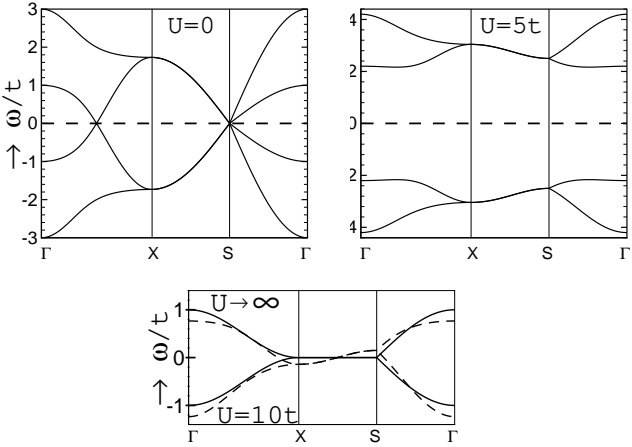}}

\vspace*{2mm}
\caption{
Electronic structure in the reduced BZ [$X=(\pi,0)$, $S=(\pi/2,\pi/2)$] as 
obtained from $H_t$ (2) for: $U=0$ (left), and $U=5t$ (right). In the bottom 
part the lower Hubbard band (centered at $\omega=0$) obtained for $U=10t$ 
(dashed lines) is compared with the $U\rightarrow\infty$ 
dispersion $\varepsilon^0_{\bf k}(\phi=0)$ (full lines).} 
\label{Fig:disp}
\end{figure}

{\it LDA+U bands\/} $-$ 
The Hamiltonian $H_t$ gives a metallic system with a bandwidth of $6t$, and 
can be seen as a tight-binding representation of the $e_g$ bands in an 
idealized structure without lattice distortions. However, LaMnO$_3$ is 
a Mott-Hubbard system with large on-site Coulomb repulsion $U$ which acts 
between the {\it occupied\/} and {\it empty\/} $e_g$ states, 
$H_U=U\sum_i f_{i0}^{\dagger}f_{i0}^{}f_{i1}^{\dagger}f_{i1}^{}$. In the 
orbital-ordered state at $n=1$ the charge fluctuations are suppressed and 
one may use the constraint $f_{i0}^{\dagger}f_{i0}^{}=1$. The Coulomb 
interaction acts then as a local potential on the unoccupied states, and the 
band structure is split into a lower and upper Hubbard band separated by 
a gap $\propto U$, as is typical for a  Mott-Hubbard insulator (Fig. 
\ref{Fig:disp}). For the manganites $U/t\approx 10$, and 
one finds that the lower Hubbard band has a dispersion $\sim 2t$, quite 
close to the $U=\infty$ limit. This resembles the strong redistribution of 
the bands in the LDA+U approach with respect to those found within the LDA, 
as obtained for LaMnO$_3$ \cite{Sat96} when the local electron-electron 
interaction terms $\propto U$ are treated in the mean-field approximation 
\cite{notegap}.

\begin{figure}
      \epsfxsize=70mm
      \centerline{\epsffile{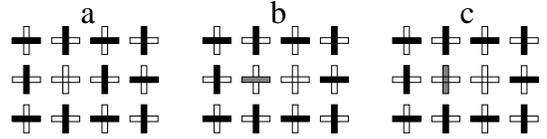}}

\vspace*{2mm}
\caption{
Schematic representation of a hole in an orbital-ordered ground state (a).
The occupied (empty) orbitals $|\mu\rangle$ and $|\nu\rangle$ are shown as 
filled (empty) rectangles. The hole can move either without disturbing
the orbital order (b), or by creating orbital excitations (c).}
\label{Fig:hop}
\end{figure}

In our framework, however, the interpretation of Eq. (\ref{eq:ht01}) is 
quite different. On the site with the hole both orbitals are empty [Fig. 
\ref{Fig:hop}(a)]. An electron on a neighboring site can hop to either one 
of these orbitals. If the orbital order is preserved in the hopping process 
[Fig. \ref{Fig:hop}(b)], the hole propagates freely with a dispersion 
determined by the first term $\sim f_{i0}^{\dagger}f_{j0}^{}$ in $H_t$. 
Unlike in the $t$-$J$ model, the hopping couples the orthogonal states 
$|\mu\rangle$ and $|\nu\rangle$ on both sublatices. In contrast, if the 
orbital order is locally disturbed by the hopping process 
$\sim f_{i1}^{\dagger}f_{j0}^{}$ [Fig. \ref{Fig:hop}(c)], an orbital 
excitation occurs \cite{string}. In this way the propagation of the hole is 
coupled to the orbital excitations determined by $H_J$. This coupling is a 
direct consequence of the strongly correlated nature of the $e_g$ electrons 
and its repercussions on the low energy electron dynamics cannot be taken 
into account within a mean-field (or LDA+U) treatment of the correlations.

{\it Orbital $t$-$J$ model\/} $-$
Our total Hamiltonian, 
\begin{equation}
{\cal H}=H_t+H_J+H_z, 
\label{orbtj}
\end{equation}
includes, in addition to $H_J$ (1) and $H_t$ (2), a crystal-field term 
$H_z=-E_z\sum_i T^z_i$, which occurs, e.g., due to unaxial pressure. It acts 
like a magnetic field in the orbital sector which modifies the occupied 
orbitals and implies {\it different\/} basis rotations by 
$\psi_{A(B)}=\frac{\pi}{4}\mp\phi$ on $A$ and $B$ sublattice \cite{Bri99}, 
respectively, with $\sin2\phi=-E_z/4J$.

{\it Free hole dispersion\/} $-$
The hopping of the hole can be expressed by the change of the orbital 
background using the Schwinger boson representation \cite{Mar91} for 
the orbital variables which obey the constraint of no double occupancy
in the limit of $U\to\infty$: $f_{i0}^{\dagger}=b_{i0}^{}h_i^{\dagger}$ and
$f_{i1}^{\dagger}=b_{i1}^{}h_i^{\dagger}$, with $h_i^{\dagger}$ standing for 
a moving fermionic hole. In the orbital ordered state the $b_{i0}$ bosons 
are condensed, $b_{i0}\simeq 1$, which leads to a free dispersion of the 
hole (Fig. \ref{Fig:disp}),
\begin{equation}
H_h=\sum_{\bf k}\varepsilon^0_{\bf k}(\phi)h_{\bf k}^{\dagger}h_{\bf k}^{},
\label{band}
\end{equation}
where $\varepsilon^0_{\bf k}(\phi)=t(-2\sin 2\phi+1)\gamma_{\bf k}$ is  
determined by the orbital order, and   
$\gamma_{\bf k}=\case{1}{2}(\cos k_x+\cos k_y)$.

{\it Orbital excitations\/} $-$
The orbital background is described by a local constraint for $T=1/2$ 
pseudospins $b_{i0}^{\dagger}b_{i0}^{}+b_{i1}^{\dagger}b_{i1}^{}=2T$, where 
$b_{i0}^{\dagger}$ and $b_{i1}^{\dagger}$ are boson operators which refer 
to the occupied and empty state at site $i$. The orbital excitations are 
calculated by using the lowest-order expansion of the constraint around the 
orbital-ordered ground state:
$T_i^x\simeq \case{1}{2}(b_{i}^{}+b_{i}^{\dagger})$ and
$T_i^z\simeq T-b_{i}^{\dagger}b_{i}^{}$, with $b_{i}\equiv b_{i1}$.
The effective boson Hamiltonian is diagonalized by a Fourier and Bogoliubov 
transformation defined by 
$b_{\bf k}^{\dagger}=u_{\bf k}^{}\alpha_{\bf k}^{\dagger}
                    +v_{\bf k}^{}\alpha^{}_{\bf -k}$, with the coefficients
$\{u_{\bf k},v_{\bf k}\}$ given in Ref. \cite{Bri99}. One finds
\begin{equation}
H_o=\sum_{\bf k}
\omega_{\bf k}(\phi)\alpha^{\dag}_{\bf k}\alpha^{}_{\bf k},
\label{magnons}
\end{equation}
where $\omega_{\bf k}(\phi)=
3J\left[ 1 +\case{1}{3}(2\cos4\phi-1)\gamma_{\bf k}\right]^{1/2}$ is the 
orbiton dispersion. This single mode, defined in the full Brillouin zone (BZ), 
is equivalent to two branches of orbital excitations obtained in the folded 
zone in Ref. \cite{Bri99}. The orbital excitations depend sensitively on 
the orbital splitting $E_z$. At orbital degeneracy ($E_z=0$) one finds 
a maximum of $\omega_{\bf k}=3J(1+\case{1}{3}\gamma_{\bf k})^{1/2}$ at the 
$\Gamma=(0,0)$ point and a weak dispersion $\sim J$. In contrast, for 
$E_z=\pm 2J$ orbital excitations are dispersionless, with $\omega_{\bf k}=3J$.

{\it Hole-orbiton coupling\/} $-$
The remaining part of Eq. (\ref{eq:ht01}) describes the hole-orbiton 
interaction [Fig. \ref{Fig:hop}(c)],
\begin{eqnarray}
H_{ho}=
t\sum_{{\bf k},{\bf q}}h_{{\bf k}+{\bf q}}^{\dagger}h^{}_{\bf k}
  \left[M_{{\bf k},{\bf q}}^{}\alpha_{\bf q}^{} \right.
       +\left. N_{{\bf k},{\bf q}}^{}\alpha_{{\bf q}+{\bf Q}}^{}+H.c.\right],
\label{orbpol}
\end{eqnarray}
where ${\bf Q}=(\pi,\pi)$,
$M_{{\bf k},{\bf q}}=2\cos 2\phi\left(u_{\bf q}\gamma_{{\bf k}-{\bf q}}
                                     +v_{\bf q}\gamma_{\bf k}\right)$, 
$N_{{\bf k},{\bf q}}=-\sqrt{3}\left(u_{\bf q}\eta_{{\bf k}-{\bf q}}
                                   -v_{\bf q}\eta_{\bf k}\right)$,
and $\eta_{k_x,k_y}=\gamma_{k_x,k_y+\pi }$. So far we have reduced the total 
Hamiltonian (\ref{orbtj}) to an effective Hamiltonian 
${\cal H}_{eff}=H_h+H_o+H_{ho}$, linearized in the slave fermion formalism, 
and treating orbital and hole dynamics on equal footing. Let us, 
for clarity, summarize the differences to the standard $t$-$J$ model 
describing a hole in a quantum antiferromagnet:
(i) the orbital model (\ref{orbtj}) contains a free dispersion of the hole
$\varepsilon_{\bf k}^0(\phi)$ which depends strongly on $E_z$;
(ii) the orbital excitations are in general different when the momentum is 
changed by a nesting vector ${\bf Q}$ 
[$\omega_{{\bf k}+{\bf Q}}(\phi)\neq \omega_{\bf k}(\phi)$], and
their dispersion varies with $E_z$, and finally,
(iii) the hole scattering on the orbital excitations has a richer analytic
structure than in the $t$-$J$ model, with new processes 
$\propto N_{{\bf k},{\bf q}}$. A qualitatively similar case to the $t$-$J$ 
model arises at $E_z=-2J$, where the free dispersion vanishes, but the 
orbital excitations are simultaneously dispersionless, so that the orbital 
model (\ref{orbtj}) reduces to the $t$-$J^z$ model. 

We investigate the spectral function and QP properties using the 
self-consistent Born-approximation (SCBA) \cite{Kane89}, known to be 
very reliable for the single hole problem \cite{Mar91}.
Treating ${\cal H}_{eff}$ in the SCBA, we find the selfenergy
\begin{eqnarray}
\Sigma({\bf k},\omega)&=& t^2\sum_{\bf q}\left[ M^2_{{\bf k},{\bf k-q}}\ 
G({\bf k-q},\omega-\omega_{\bf q}(\phi)) \right. \nonumber \\ 
& &\left. \hskip .8cm 
+ N^2_{{\bf k},{\bf k-q}}\ G({\bf k-q},\omega-\omega_{\bf q+Q}(\phi))\right], 
\label{self}
\end{eqnarray}
which, together with the Dyson equation for the hole Green function
$G({\bf k},\omega)$, represents a closed set of equations and was solved 
self-consistently by numerical iteration on a grid using 160 ${\bf k}$-points.

\begin{figure}
      \epsfxsize=80mm
\caption{
Contour plot of the spectral function (left) and density of states DOS 
(right, blue lines) for $J/t=0.2$, $E_z=0$ (top) and $E_z=2J$ (bottom). 
Quasiparticle states at the top of the valence band are highlighted in 
yellow color, while free hole bands $\varepsilon^0_{\bf k}(\phi)$ and DOS 
are shown by red lines.}
\label{fig:spectra}
\end{figure}

{\it Quasiparticles\/} $-$
The spectral functions found in SCBA,
$A({\bf k},\omega)=-\case{1}{\pi}{\rm Im}G({\bf k},\omega)$, 
consist of a QP band close to the Fermi energy, while the excitations deep 
in the valence band are incoherent, taking typical values of $J/t<0.5$. 
In Fig. \ref{fig:spectra} we present contour plots of the spectral function. 
The total width of the spectrum is comparable 
to the free dispersion, as shown by the density of states, but the coupling 
to the orbitals leads to a strong redistribution of spectral weight. 

As the accurate orbital wave functions are not known \cite{Mur98}, we 
investigated the spectra for a few representative orbital ordered states. 
The QP part of the spectral function obtained for the alternating 
$(|x\rangle\pm|z\rangle)/\sqrt{2}$ orbital order at $E_z=0$ resembles the 
free dispersion with a maximum at the $\Gamma$ point, but its bandwidth is 
reduced to $\sim J$. When $E_z=2J$, which corresponds to the alternation of 
$3x^2-r^2/3y^2-r^2$ orbitals \cite{notee}, the spectral function changes 
markedly (Fig. 
\ref{fig:spectra}). The incoherent weight is now distributed more smoothly 
and the original free dispersion $\varepsilon^0_{\bf k}(\phi)$ (red lines) 
can still be recognized in the spectral function but is strongly damped. 
Most strikingly, there is only one QP band with appreciable weight left in 
a limited region of the BZ, while the second band is absorbed by the 
continuum. In the opposite case of $E_z=-2J$, i.e., $x^2-z^2$/$y^2-z^2$ 
orbitals alternate, the spectrum (not shown) is identical to the {\it ladder 
spectrum\/} of the $t$-$J^z$ model; it is ${\bf k}$-independent and consists 
of a set of $\delta$-functions at approximately equal energy intervals 
\cite{Kane89}. In this respect the spectrum found at $E_z=0$ can be viewed 
as a compromise between the extremes of the ladder spectrum, from which it 
retains some character of the enhanced/reduced spectral weight at regular 
energy intervals, and the $E_z=2J$ spectrum in which the dispersive features 
are smeared out over the entire band.

\begin{figure}
      \epsfxsize=80mm
      \epsffile{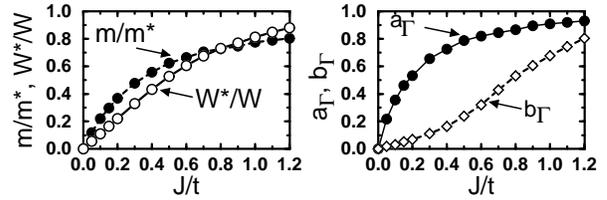}
\vspace*{2mm}
\caption{
Quasiparticle properties at $E_z=0$ as functions of $J/t$:
 (a) inverse effective mass $m/m^*$ (full circles) and the bandwidth 
     $W^*/W$ (empty circles) in units of $m$ and $W$ defined by
     $\varepsilon_{\bf k}^0(0)$ (\protect\ref{band}); 
 (b) weights of high-energy QP $a_{\Gamma}$ (filled circles) and low-energy 
     QP $b_{\Gamma}$ (open squares) at the $\Gamma$ point.}
\label{fig:mstar}
\end{figure}

Finally, we analyze the QP properties at orbital degeneracy ($E_z=0$). 
Two QP states at the $\Gamma$ point have the spectral weights $a_{\Gamma}$ 
and $b_{\Gamma}$, respectively, and determine the bandwidth $W^*$ of the QP 
band (Fig. \ref{fig:mstar}). At low doping the states at the top of the QP 
band are filled by holes, and thus the transport properties depend on the 
effective mass found at the $\Gamma$ point, 
$(\partial^2 E_{\bf k}/\partial k^2)|_{{\bf k}=0}\propto m/m^*$, where 
$E_{\bf k}$ is the QP energy. In the weak-coupling regime ($J>t/2$) the QP 
bandmass is almost unrenormalized, while it is strongly enhanced in the 
strong-coupling regime ($J<t/2$). For the manganites $J/t\simeq 0.1$ 
\cite{noteratio}, so that $W^*/W\simeq 0.1$, and the QP mass $m^*$ is 
{\it increased by an order of magnitude\/} due to the dressing of a hole by 
the orbital excitations. The weights of the two QP branches at the $\Gamma$ 
point are strongly reduced in the strong-coupling regime, but the QP at the 
top of the valence band is still quite distinct. 

{\it Summary\/} $-$
In conclusion, we have shown that a hole moving in FM planes of LaMnO$_3$ 
couples strongly to orbital excitations which results in a large 
redistribution of spectral weight compared to single-electron or mean-field 
treatments. The coherent quasiparticle band and the incoherent part of the 
spectral function depend critically on the orbital ordering and would 
therefore be {\it different\/} for FM planes of cubic and layered manganites. 
This prediction could be verified by angle resolved photoemission.
Unfortunately, such experiments have not yet been performed in the 
orbital-ordered phase at low doping concentration. Recent experiments 
by Dessau {\it et al.} \cite{Des98} on the highly doped layered system 
La$_{1.2}$Sr$_{1.8}$Mn$_2$O$_7$ showed strong incoherent features which 
were discussed in terms of small lattice polarons. 
Our study points out that the incoherence observed in these experiments
might be attributed instead to orbital excitations.

One of us (J.v.d.B.) acknowledges with appreciation the support by the 
Alexander von Humboldt-Stiftung, Germany. A.M.O. thanks the Committee of
Scientific Research (KBN) Project No.~2 P03B 175 14 for support.




\begin{references}
\bibitem{noteratio} Taking realistic parameters of LaMnO$_3$, one finds
                    (the effective) $U\simeq 3.8$ eV for $|^6A_1\rangle$
                    state which with $t=0.4$ eV gives $J/t=t/U\simeq 0.1$ 
                    \protect\cite{Fei99}. The same form of the effective 
                    interaction follows also from the cooperative JT effect 
                    which would enhance $J$ by a factor close to two.
\end{references}
\end{document}